\documentclass[pra,reprint,superscriptaddress]{revtex4-2}

\usepackage{amsmath}
\usepackage{mathtools}
\usepackage[T1]{fontenc}
\usepackage[dvipsnames]{xcolor}
\definecolor{dark-blue}{rgb}{0,0.2,0.6}
\usepackage[colorlinks,%
            pdfusetitle,%
            urlcolor=dark-blue,%
            citecolor=dark-blue,%
            linkcolor=dark-blue]{hyperref}
\usepackage{graphicx}
\usepackage{comment}
\usepackage{newtx}

\DeclareMathAlphabet{\mathcal}{OMS}{cmsy}{m}{n}
\DeclareSymbolFont{CMAlt}{OMX}{cmex}{m}{n}
\DeclareMathSymbol{\sumop}{\mathop}{CMAlt}{"50}
\DeclareMathSymbol{\intop}{\mathop}{CMAlt}{"52}

\newcommand{\figu}[1]           
{Fig.~\ref{#1}}
\newcommand{\secu}[1]           
{Sec.~\ref{#1}}

\newcommand{\up}{\ensuremath{\uparrow}}
\newcommand{\dw}{\ensuremath{\downarrow}}
\newcommand{\ket}[1]{\ensuremath{|{#1}\rangle}}
\newcommand{\bra}[1]{\ensuremath{\left\langle{#1}\right|}}

\newcommand{\defeq}{\stackrel{\text{def}}{=}}

\begin{document}

\author{A.~Amaricci}
\affiliation{CNR-IOM -- Istituto Officina dei Materiali, Consiglio Nazionale delle Ricerche, 34136 Trieste, Italy}
\author{A.~Richaud}
\affiliation{Scuola Internazionale Superiore di Studi Avanzati (SISSA), 34136 Trieste, Italy}
\affiliation{Departament de F\'isica, Universitat Polit\'ecnica de Catalunya, 08034 Barcelona, Spain}
\author{M.~Capone}
\affiliation{Scuola Internazionale Superiore di Studi Avanzati (SISSA), 34136 Trieste, Italy}
\affiliation{CNR-IOM -- Istituto Officina dei Materiali, Consiglio Nazionale delle Ricerche, 34136 Trieste, Italy}
\author{N.~\surname{Darkwah Oppong}}
\email[E-mail address: ]{ndo@caltech.edu}
\affiliation{California Institute of Technology, Pasadena, CA 91125, USA}
\author{F.~Scazza}
\email[E-mail address: ]{francesco.scazza@units.it}
\affiliation{Department of Physics, University of Trieste, 34127 Trieste, Italy}
\affiliation{CNR-INO -- Istituto Nazionale di Ottica, Consiglio Nazionale delle Ricerche, 34149 Trieste, Italy}

\title{Engineering the Kondo impurity problem with alkaline-earth atom arrays}

\begin{abstract}
We propose quantum simulation experiments of the Kondo impurity problem using cold alkaline-earth(-like) atoms (AEAs) in a combination of optical lattice and optical tweezer potentials. 
Within an \textit{ab initio} model for atomic interactions in the optical potentials, we analyze hallmark signatures of the Kondo effect in a variety of observables accessible in cold-atom quantum simulators. 
We identify additional terms not part of the textbook Kondo problem, mostly ignored in previous works and giving rise to a direct competition between spin and charge correlations---strongly suppressing Kondo physics.
We show that the Kondo effect can be restored by locally adjusting the chemical potential on the impurity site, and
we identify realistic parameter regimes and preparation protocols suited to current experiments with AEA arrays. 
Our work paves the way for novel quantum simulations of the Kondo problem and offers new insights into Kondo physics in unconventional regimes.
\end{abstract}

\maketitle

\section{Introduction}
The Kondo effect (KE) is a cornerstone of modern condensed matter physics. First discovered as a low-temperature resistivity minimum in metals \cite{DeHaas1934}, it was named after Jun Kondo, who provided the first explanation of the phenomenon in terms of his celebrated model of a single magnetic impurity embedded in a bath \cite{Kondo1964}. The quest for a complete understanding of the KE has been the driving force for many advances in quantum many-body theories, culminating in the development of the numerical renormalization group \cite{Wilson1975}.
The KE arises from the formation, below the Kondo temperature $T_K$, of a singlet between the impurity and the bath, which quenches the entropy of the impurity magnetic moment---shaping the low-energy behavior of the system resulting in striking thermodynamic and transport properties~\cite{Hewson1993,Kondo2012}.
Fingerprints of the KE are ubiquitous in solid-state systems, ranging from heavy-fermion compounds~\cite{Hewson1993,Coleman2007,Si2010,Song2022PRL} to quantum dots~\cite{Goldhaber-Gordon1998,Cronenwett1998,Jeong2001,Grobis2007}.
Kondo-related problems remain at the forefront of condensed-matter research, especially in extensions featuring multiple reservoirs \cite{Potok2007,Lopes2020} or impurities, as in the so-called Kondo lattice \cite{Doniach1977,Coleman2007}. This owes to the rich interplay between Kondo screening and fermion-mediated interactions between impurities~\cite{Hewson1993,Coleman2001,Lohneysen2007,Si2010,Pruser2014}.
Moreover, dynamical mean-field theory, which maps a solid onto a self-consistent impurity, has provided a new perspective connecting the KE with the Mott transition~\cite{Georges1992}.

New opportunities to study the physics of the KE are offered by quantum simulations with cold atomic systems~\cite{Jaksch2005}, allowing to target regimes that are challenging for conventional solid-state systems and theoretical approaches, especially concerning the investigation of real-time \cite{Nordlander1999,Kaminski2000,Lobaskin2005,Anders2005,Weichselbaum2009,Ashida2018a,KanaszNagy2018} and real-space dynamics~\cite{Holzner2009,Affleck2010,Busser2010,Mitchell2011,Park2013,Borzenets2020}.
In this context, alkaline-earth(-like) atoms (AEAs) in state-dependent optical lattices emerged as a promising platform for cold-atomic implementations of Kondo problems~\cite{Gorshkov2010,Foss-Feig2010,Foss-Feig2010a,Nakagawa2015,Zhang2016,KanaszNagy2018,Goto2019}.
Here, AEAs in the ${}^1$S$_0$ ground and ${}^3$P$_0$ metastable state, can mimic electrons in the conduction band and localized magnetic moments, respectively.
Pioneering experiments with ytterbium atoms have carefully characterized the relevant interactions and demonstrated the key building blocks for quantum simulation experiments of the Kondo model~\cite{Scazza2014,Cappellini2014,Riegger2018,Ono2019,Abeln2021,Bettermann2020,Ono2021}.
However, to date, experimental observations of the cold-atomic KE are still lacking.
This can be attributed to several experimental challenges which include reaching sufficiently low temperatures, strong dissipation from collisions of atoms in the metastable state~\cite{Scazza2014,Riegger2018,Sponselee2019,Ono2021}, and the unavoidable presence of parasitic terms not found in the ``plain vanilla'' Kondo model. 
Recent advancements of atom-array experiments and their pairing with conventional optical-lattice systems~\cite{Weitenberg2011,Koepsell2019,Yan2022,Spar2022,Young2024} have made a range of powerful new experimental tools available. 
In particular, programmable optical tweezers could be harnessed to compensate for parasitic terms at the level of single lattice sites~~\cite{Weitenberg2011,Koepsell2019,Young2022,Lebrat2024,Bohrdt2024} and enable the experimental observation of the KE in cold atomic systems.

In this work, we focus on the single-impurity limit of the two-orbital Hubbard Hamiltonian that accurately describes fermionic AEAs in optical lattices~\cite{Gorshkov2010}.
Exploiting state-selective optical traps, one of the two orbitals can be localized, effectively pinning an impurity into a bath of mobile fermions~\cite{Riegger2018,Ono2021}.
In such setting, spin-exchange and density-density onsite interactions between mobile fermions and the localized impurity have comparable strengths, owing to the distinct inter-orbital singlet and triplet channels contributing to the low-energy atomic scattering problem~\cite{Gorshkov2010,Scazza2014,Zhang2020}.
We find that repulsive impurity-bath interactions hinder the buildup of spin correlations---even for relatively small strengths where the occupation of the impurity site with mobile fermions is not strongly suppressed.
Leveraging site-resolved light shifts, we establish a simple protocol to lift this limitation and recover the KE at temperatures comparable to those reached in state-of-the-art cold-atom experiments~\cite{Taie2022,Xu2025}.
We study several observables suitable to experimental measurements, demonstrating that small-scale AEA arrays provide access to Kondo physics in the unconventional regime of strong exchange.

We note that previous theoretical studies of the Kondo impurity problem in cold atomic systems have largely neglected detrimental spin-independent interaction terms~\cite{Nakagawa2015,Zhang2016,Goto2019}. Owing to their effect, in the strongly interacting regime and without applying corrective light shifts, a weak Kondo exchange arises only at second perturbative order through off-resonant fermion tunneling onto the impurity site---i.e., through superexchange processes~\cite{KanaszNagy2018,Riegger2018,Stefanini2024}. Our work informs new cold-atom implementations of the Kondo problem without relying on superexchange couplings, exploiting  recently demonstrated capabilities in combining optical tweezers and lattices.

\smallskip

The rest of this paper is organized as follows. In the forthcoming \secu{SecModel} we present the main model used in this work, including the atomic-physics origin of the parameters. The model derivation and further insight into the numerical methods are presented in the Appendices~\ref{Sec:DerivModel} and~\ref{Sec:NumericalMethods}.
In Sec.~\ref{SecSignatures} we discuss the observable signatures of Kondo physics and analyze the effects of the local inter-orbital interaction and impurity tunneling. 
Based on these results, in Sec.~\ref{SecProtocol} we propose a detailed experimental protocol to engineer Kondo physics with AEA arrays. Finally, in Sec.~\ref{SecConclusions} we summarize our results and discuss some future perspectives of our work.

\begin{figure}[t!]
\includegraphics[width=\linewidth]{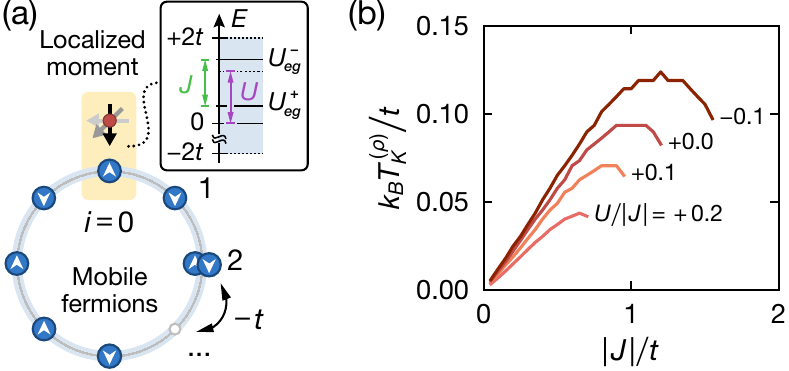}
    \caption{\label{fig1}%
        \textbf{Exploring Kondo physics in AEA arrays.}
        (a)~Illustration of the cold-atomic Kondo impurity problem studied in this work.
        Mobile fermionic atoms in two spin states (blue circles with arrows) hop with amplitude~$t$ on a chain with periodic boundary conditions.
        At the impurity site ($i=0$), a pinned impurity atom serves as localized moment (red circle with arrow).
        The spin-exchange coupling~$J$ with the mobile fermions and the potential $U$ are set by the atomic singlet and triplet interaction strengths~$U_{eg}^\pm$ (see main text).
        (b)~Kondo temperature~$\smash{T_K^{(\rho)}}$ 
        for variable~$J/t$ and~$U/|J|$ (colored lines).
        Termination of lines indicates that the resistivity minimum ceases to exist for the corresponding parameters.
        }
\end{figure}

\bigskip
\section{Kondo model and tuning parameters}\label{SecModel}
We consider a one-dimensional (1D) tight-binding model of cold fermionic AEAs with spin states~\mbox{$\sigma\in\{\downarrow, \uparrow\}$} such as~${{}^\text{171}\text{Yb}}$. 
Treating atomic interactions in an optical lattice \emph{ab initio} and adding a suitable localizing potential for the impurity, the full two-orbital Hubbard model can be reduced to a Kondo-like Hamiltonian. A complete derivation is presented in Appendix~\ref{Sec:DerivModel}. 
This describes mobile fermions in a half-filled band (width~$4t$) hopping between $L$~sites and interacting with a localized magnetic moment at a given impurity site (\mbox{$i=0$}) [see Fig.~\ref{fig1}(a)]: 

\begin{equation}\label{eq:ham}
    \hat{\mathcal{H}} = - t\!\sum_{\langle ij \rangle,\sigma}\! \left( \hat{c}^\dagger_{i\sigma} \hat{c}^{\vphantom{\dagger}}_{j\sigma} + \mathrm{h.c.} \right) - J \,\hat{\mathbf{S}}_0 \cdot \hat{\mathbf{s}}_0 + U\sum_{\sigma} \hat{n}_{0\sigma}.
\end{equation}
Here, $\langle i j\rangle$ denotes nearest-neighbor lattice sites, $\hat{c}_{i\sigma}^{\dagger}$ ($\hat{c}_{i\sigma}^{\vphantom{\dagger}}$) creates (annihilates) a mobile fermion on lattice site~$i$ with spin $\sigma$, and $\hat{n}_{i\sigma} = \hat{c}_{i\sigma}^{\dagger}\hat{c}_{i\sigma}^{\vphantom{\dagger}}$, while $\hat{\mathbf{S}}_0$ (referred to as $\hat{\mathbf{S}}$ in the following) and $\hat{\mathbf{s}}_i$ correspond to the pseudo-spin operators of the localized moment and mobile fermions, respectively.
The (anti-ferromagnetic) spin-exchange energy~$J<0$, as well as the central-site potential~$U$ are effective tuning parameters of the model. 
In particular, the Kondo coupling $J$ and the local potential $U$ emerge from the decomposition of the inter-orbital interaction term within a two-orbital Hubbard model~\cite{Gorshkov2010}, assuming complete localization of the impurity fermion. The effective value of $U$ can be tuned through a local potential $\mu$ of the $\ket{g}$-state at the impurity site $i=0$ (see Appendix~\ref{Sec:DerivModel}).  
We investigate the finite-temperature physics by performing exact diagonalization of both $\hat{\mathcal{H}}$ and the full two-orbital model with periodic boundary conditions for~$L=7$ (see Appendix~\ref{Sec:NumericalMethods}). 

In experiments with fermionic AEAs~\cite{Riegger2018,Ono2021}, the localized moment typically corresponds to an atom in the metastable clock state ${}^3$P$_0$ (denoted by $\ket{e}$), whereas the mobile fermions correspond to atoms in the electronic ground state ${}^1$S$_0$ (denoted by $\ket{g}$).
The coupling parameters of Eq.~\eqref{eq:ham} are directly connected to the two available on-site inter-orbital pair states, $\ket{eg^\pm} = \frac{1}{2} \left(\ket{ge} \pm \ket{eg} \right) \otimes \left( \ket{\uparrow\downarrow} \mp \ket{\downarrow\uparrow} \right)$ with energy $U_{eg}^\pm$, via $J = U_{eg}^+ - U_{eg}^-$ and $U = \frac{1}{4} U_{eg}^+ + \frac{3}{4} U_{eg}^-$ [Fig.~\ref{fig1}(a)].
Therefore, $U$ and $J$ are fully determined by atomic scattering properties (see Appendix~\ref{Sec:DerivModel}).
However, we first treat them as freely tuneable parameters to identify optimal regimes for observing the KE.
To this end, we estimate the Kondo temperature~$T_K$ from a linear-response numerical calculation  of the well-known resistivity minimum for variable~$J<0$ and~$U$ [Fig.~\ref{fig1}(b)]. Details are provided in Appendix~\ref{Sec:NumericalMethods}.  
Accordingly, we set $\smash{T_K^{(\rho)}}$ as the temperature at which the resistivity $\rho(T)$ has a local minimum (if any exists).  

\begin{figure*}
    \includegraphics[width=\linewidth]{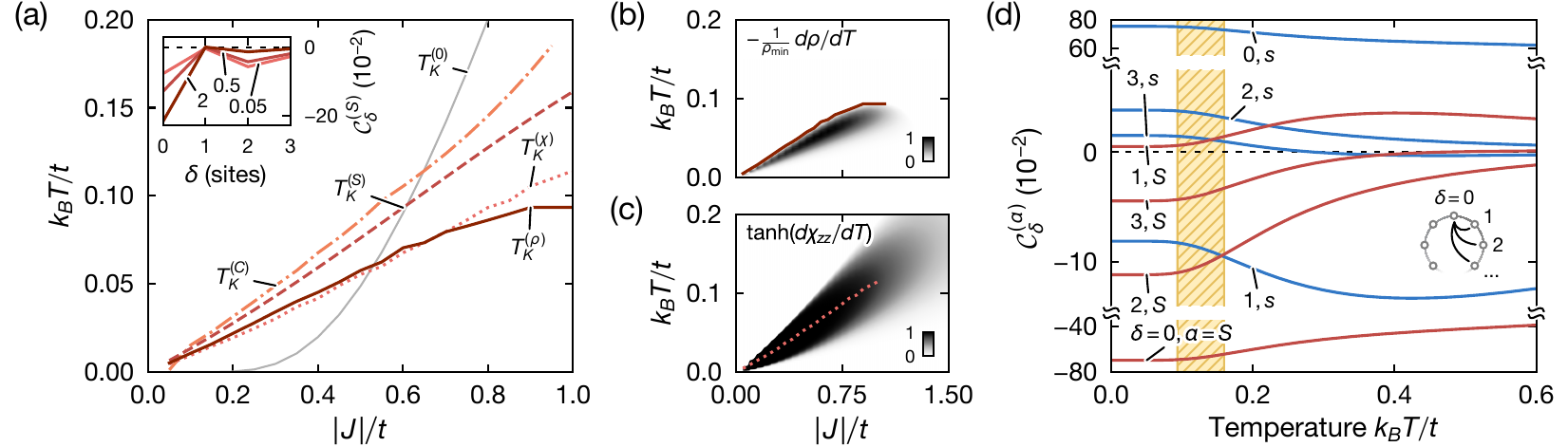}
    \caption{\label{fig2}%
        \textbf{Emergence of the Kondo effect in a small-scale system.} 
        (a)~Estimated Kondo temperature scales in comparison with the perturbative exponential behavior (gray line).
        See main text for the definitions of the different symbols. Inset:~Impurity-fermion spin correlations $\smash{\mathcal{C}^{(S)}_{\delta}}$ at $T=0$ for different values of $|J|$ indicated in the labels.
        (b,c)~Charge and spin physics of the KE as a function of the exchange coupling $J$ and temperature $T$. The dark areas delimit the regions of Kondo crossover regime. 
        The colors scales correspond to the normalized temperature derivative of (b)~the resistivity and (c)~spin susceptibility. 
        Red lines denote $\smash{T_K^{(\rho)}}$ and $\smash{T_K^{(\chi)}}$ obtained for Eq.~\eqref{eq:ham} with $U=0$.
        (d)~Spin correlations $\smash{\mathcal{C}^{(\alpha)}_{\delta}}$ for $|J|/t=1$ at variable~$T$ and distances~$\delta$ from the impurity  (inset).
        The shaded yellow region denotes $\smash{T_K^{(\rho)} < T < T_K^{(S)}}$.
    }
\end{figure*}

We first focus on the case~$U = 0$, i.e., the s-d model, which captures the essential physics of the KE~\cite{Hewson1993}. 
While solid-state systems generally feature small spin-exchange interactions~$|J| \ll t$~\cite{Hewson1993}, cold atomic systems can realize much larger values~$|J|\gg t$~\cite{Scazza2014,Goban2018,Ono2019,Bettermann2020}.
Previous studies have indicated that this can significantly increase the Kondo temperature~$T_K$~\cite{Zhang2016,Goto2019}.
However, our results suggest that $T_K^{(\rho)}$ only increases until~$|J| \approx t$ and the resistivity minimum quickly ceases to exist once~$|J| > t$~[Fig.~\ref{fig1}(b)].
This can be qualitatively understood by considering that in 1D, the ground-state singlet formed by impurity and mobile fermion localizes for~$|J|\gg t$, suppressing transport through the impurity site.%

For finite~$U$, the physics strongly depends on the sign of this additional tuning parameter.
While small negative values lead to a modest increase of $T_K^{(\rho)}$, the resistivity minimum vanishes for~$-U/J \gtrsim 0.2$.
This can be explained by considering that~$U$ acts as an effective well ($U<0$) or repulsive barrier ($U>0$) localized at the impurity site, determining its occupation~$\langle\hat{n}_{0\sigma}\rangle$.
In this way, Kondo physics arising from spin-exchange between impurity and mobile fermions, can be enhanced ($-t \lesssim U < 0$) or strongly suppressed ($U > 0$).
Our numerical results show that~$U/|J| \lesssim 0$ and~$|J|\approx t$ are optimal for maximizing the Kondo temperature $T_K^{(\rho)}$.
For previous cold-atom experiments with $U\gg t$ and $J > t$~\cite{Riegger2018,Ono2021}, our results suggest that the KE is suppressed. %

\section{Signatures of Kondo physics}\label{SecSignatures}

So far, we have considered only the resistivity minimum as an indicator for the KE.
In order to corroborate the emergence of a single Kondo energy scale characterizing the crossover in our small-scale system, we identify other observables carrying signatures of the KE in both charge and spin degrees of freedom.
The buildup of Kondo screening can be captured by the temperature evolution of the impurity spin-spin susceptibility~$\chi_{zz}(T)$:
the high-temperature Curie-like behavior~($\chi_{zz}\propto 1/T$) associated with a nearly free local moment turns into a Pauli behavior corresponding to a large effective mass,~$\chi_{zz}(T\to0)\propto m^*$ (see Appendix~\ref{Sec:NumericalMethods}). 
We identify $T_K^{(\chi)}$ finding the inflection point $\max_T d\chi_{zz}/dT$ separating the two regimes. 
The evolution from a local Fermi-liquid to a decoupled impurity state is also signaled by a maximum in the specific heat~$C_v$ as a function of $T$ (Schottky anomaly), whose position defines $T_K^{(C)}$.
Correspondingly, we define $\smash{T_K^{(S)}}$ as the temperature for which the impurity entropy reaches $S=k_B \ln{2}$, signaling that it is locked in a singlet state.
Further, the formation of a many-body singlet gives rise to significant off-site correlations at distance~$\delta$ from the impurity site, $\mathcal{C}^{(S)}_{\delta}=\langle \hat{S}_{z} \,\hat{s}_{z\delta}\rangle$, as well as $\mathcal{C}^{(s)}_{\delta}=\langle \hat{s}_{z0} \,\hat{s}_{z\delta}\rangle$ connecting the spin of distant fermions.

Figure~\ref{fig2} displays an overview of numerical results for $U = 0$. 
The trend of the Kondo energy scale $k_BT_K$, estimated by the markers introduced above, 
exposes the peculiarity of our small-scale system.
As visible in Fig.~\ref{fig2}(a), all Kondo temperature estimates display a quasi-linear evolution as a function of~$|J|$, which is contrasted against the textbook exponential behavior~\cite{Hewson1993}. 
Tuning the coupling $|J|\sim t$, the Kondo temperature reaches values of about $k_B T_K/t \sim 0.1$, comparable to temperatures achieved in state-of-the-art quantum microscope experiments~\cite{Taie2022,Xu2025}.

The zero-temperature impurity-fermion correlator~$\mathcal{C}^{(S)}_{\delta}$ offers an insight on how the KE can emerge at strong coupling---even in a small system [Fig.~\ref{fig2}(a) inset]. %
For~$|J|\sim t$, the Kondo screening cloud spreads over few sites, while 
off-site correlations become more pronounced as~$|J|/t$ is decreased.
Figure~\ref{fig2}(b,c) shows the $T$ and $J$ dependence of resistivity and impurity spin-spin susceptibility, capturing the Kondo crossover in charge and spin channels, respectively.

Next, we explore the onset of the KE in varying-range spin-spin correlations [Fig.~\ref{fig2}(d)].
The on-site correlator $\mathcal{C}^{(S)}_{\delta=0}<0$ evidences the formation of a robust local singlet state upon entering the Kondo regime.
Remarkably, all other correlations show a noticeable change of amplitude with respect to the their high-temperature counterparts, offering a potential route to detect signatures of Kondo physics already at temperatures of the order of $k_BT/t\sim 1$ in experiments. 
Such correlations may be probed through single-atom microscopy, allowing to retrieve the spatially-resolved impurity and fermions spin in real time, thus mapping the spatio-temporal character of the Kondo screening cloud~\cite{KanaszNagy2018,Borzenets2020}.

\section{Realizing the Kondo problem with ytterbium atom arrays}\label{SecProtocol}
We have shown that~$|J| \approx t$ and~$U \simeq 0$ maximize the Kondo temperature for the mesoscopic systems sizes studied here.
Here, we discuss an experimental setup tailored to engineer such parameter regime.
Our proposed implementation leverages the atomic species ${}^{171}$Yb ($I=1/2$), which is characterized by a favorable set of atomic properties.
It does not only feature vanishing density-density interactions in the ground state, i.e., between mobile fermions, but also exhibits negative inter-orbital spin-exchange interaction~$J<0$~\cite{Ono2019,Bettermann2020}.
However, the atomic properties of ${}^{171}$Yb yield~$U_{eg}^\pm > 0$ resulting in~$U \simeq 2.75\,|J|$, which strongly suppresses the KE.

\begin{figure}[t]
\includegraphics[width=\linewidth]{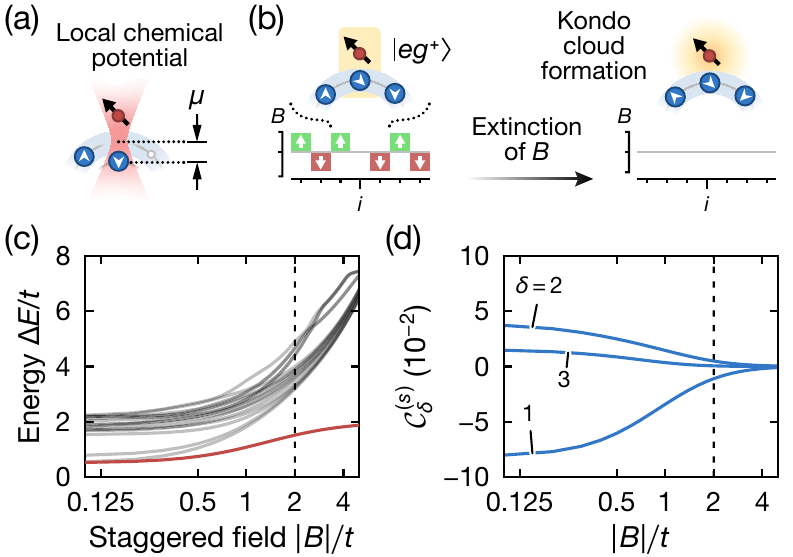}
\caption{\label{fig3}
    \textbf{Adiabatic ground-state preparation of a small-scale atomic Kondo problem.}
    (a) Illustration of AEA atoms trapped in a combined optical potential featuring a state-selective lattice for $g$-state particles and a state-dependent tweezer at site $i=0$, enabling to pin the impurity and precisely adjust the local onsite potential $\mu$.
    (b) Illustration of the staggered magnetic field $B$ and ground states for the high-$B$ (left) and low-$B$ (right) limits. The state $\ket{\kappa}$ sketched on the left can be prepared experimentally with high fidelity.
    (c) Energies of the lowest $M=20$ excitations with respect to the ground state, as a function of a variable staggering field strength~$B$.
    The energy $\Delta E_1$ of the lowest excitation (red line) remains $\Delta E_1\sim t$ from $|B|/t \gg 1$ to $|B|/t \ll 1$. %
    (d) Bath-spin correlators $\smash{\mathcal{C}^{(s)}_{\delta}}$ for $\delta = 1, 2, 3$, showing the build-up of nonlocal fermion correlations mediated by the impurity while the staggered field is lowered.
}
\end{figure}

We now briefly discuss how a state-dependent optical tweezer can be utilized to address this challenge and restore the KE at elevated temperatures, as sketched in Fig.~\ref{fig3}(a).
Without loss of generality, we assume both impurity and mobile fermions are trapped in an optical lattice~\cite{Riegger2018,Ono2021}.
The optical tweezer focused onto the impurity site introduces a local chemical potential term~$\mu (\hat{n}_{0\downarrow} + \hat{n}_{0\uparrow})$ which compensates the last term in Eq.~\eqref{eq:ham} when~$\mu = - U$.
Choosing an optical lattice wavelength at the clock-state tune-out wavelength~\cite{Hoehn2024} and an attractive optical tweezer wavelength close to the ground-state tune-out wavelength~\cite{Hoehn2023}, the impurity does not see a periodic lattice potential and can be fully localized by the tweezer potential as in Eq.~\eqref{eq:ham}. 
In this way, our approach also resolves the challenge concerning the residual mobility of the impurity in a typical state-dependent optical lattice~\cite{Riegger2018,Ono2021}.

While the system studied here could be prepared with conventional evaporative cooling techniques in cold-atomic experiments, 
an alternative route is the preparation from an initial $(\hat{S}_z \otimes \hat{n}_{i\sigma})$-product state adiabatically connected to the many-body ground state~\cite{Spar2022,Yan2022,Omran2019S}.
Such a product state can be prepared with a bottom-up approach, as recently demonstrated~\cite{Young2022,Young2024}.
For $L=7$, we carry out a state analysis to identify the relevant Fock states contributing to the KE. 
For $|J| \approx t$, we find that two specific zero-magnetization configurations, $\ket{\psi}={|\!\!\up\rangle_S} \otimes{|\!\up\dw\up\dw\dw\up\dw\rangle_s}$ and $\ket{\vartheta} = {|\!\dw\rangle_S} \otimes{|\!\up\dw\up\up\dw\up\dw\rangle_s}$, exhibit the largest contribution to the Kondo singlet. 
Any other spin configuration can be penalized with a tunable energy gap by adding a staggered, longitudinal effective magnetic field $B$ acting only on the bath sites, $i\neq0$ [Fig.~\ref{fig3}(b)]. 
This could be experimentally realized through optical site-selective and spin-dependent vector light shifts \cite{Hui2021,Bohrdt2024}, which can be implemented using optical tweezers operated close to the intercombination transition~\cite{Taie2010,Pagano2019}. In the large-field limit, $|B|\gg t$, the superposition $\ket{\kappa}=(\ket{\psi}-\ket{\vartheta})/\sqrt{2}$ featuring a local singlet pair at the impurity site matches asymptotically the ground state of the system. 
The state $\ket{\kappa}$ could be experimentally prepared by:
(i)~creating a unity-filled array of $\ket{g\!\up}$ atoms;
(ii)~adding a $\ket{g\!\dw}$ atom to the central (impurity) site;
(iii)~optically driving this pair to the $\ket{eg^+}$ state; 
(iv) flipping a subset of bath atoms to $\ket{g\!\dw}$, as targeted.
The latter steps require high-fidelity control over the nuclear spin and clock state of individual atoms, as recently demonstrated in AEAs array platforms for quantum information processing~\cite{Jenkins2022,Ma2022,Norcia2023,Lis2023}. 

To evaluate the feasibility of reaching the many-body ground state from $\ket{\kappa}$, we calculate the spectrum of the system for $|J|=t$ as a function of the staggering field amplitude~$|B|$.
In Fig.~\ref{fig3}(c), the excitation gaps~$\Delta E$ from the ground state are shown as a function of $|B|$.
For decreasing $|B|$, the smallest excitation gap never vanishes and settles to $\Delta E_1 \sim t$ for $B\ll t$. 
Starting at $|B| \gg t$, an annealing protocol can thus be followed to maintain the system in the ground state while $B$ is adiabatically reduced, mapping the onsite spin-singlet pair $\ket{eg^+}$ to a non-local Kondo singlet.
This is confirmed by the amplitude of the bath-spin correlations $\mathcal{C}^{(s)}_{\delta}$ [Fig.~\ref{fig3}(d)], which become significant while the annealing field is decreased---a feature that could be experimentally measured by single-atom resolved spin readout.
Further, $\chi_{zz}$ could be extracted by measuring the impurity magnetization as a function of a real (non-staggered) external magnetic field, acting as a relative energy shift $\Delta_{eg}$ between states $\ket{e\!\up, g\!\dw}$ and $\ket{e\!\dw, g\!\up}$ \cite{Scazza2014} at the impurity site. %
An external magnetic field has indeed a similar effect on the KE as the temperature---it causes the localized moments to become free in the limit $\Delta_{eg} \gg |J|$, just as for $T\gg T_K$~\cite{Coleman2007}.

\smallskip
\smallskip

\section{Emergent Kondo Lattice Behavior}\label{SecKondoLattice}
\begin{figure}
  \includegraphics[width=\linewidth]{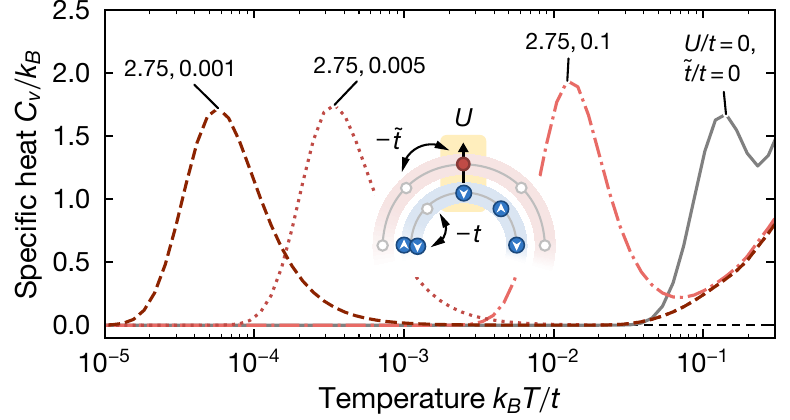}
  \caption{\label{fig4}
    \textbf{Kondo lattice effects in a single-impurity system with finite impurity tunneling.}
    Specific heat $C_v$ as a function of the
    temperature for $J=-t$, $U=2.75\,t$ and various values of $\tilde{t}$. The Schottky anomaly of the standard Kondo model
    (solid gray line) is shown for reference. The sketch illustrates the mobile-impurity Kondo problem in Eq.~\eqref{eq:ham2}, as implemented with AEAs in a state-dependent 1D optical lattice.  
  }
\end{figure}

In previous studies~\cite{Gorshkov2010,Riegger2018,KanaszNagy2018,Ono2021}, state-dependent lattices have been considered to engineer distinct mobilities for states $\ket{g}$ and $\ket{e}$.
In experiments, however, it remains challenging to fully localize the impurity \cite{Riegger2018,Ono2021} and reduce the $\ket{e}$-state tunneling~$\tilde{t}$ such that it becomes only relevant for time scales beyond those associated to the impurity scattering, $1/\tilde{t} \gg 1/|J|$.
A significant tunneling of the impurity over the lattice has not been previously taken into account; we study here its impact on the formation of a Kondo resonance, considering the two-orbital Hamiltonian

\begin{equation}\label{eq:ham2}
\begin{split}
    \hat{\mathcal{H}}^\prime = \hat{\mathcal{H}}  &-\tilde{t} \!\!\sum_{\langle ij \rangle,\sigma} 
    \left(\hat{d}^\dagger_{i\!\sigma} \hat{d}^{\vphantom{\dagger}}_{j\!\sigma} \!+ \mathrm{h.c.} \right)\\
    & - J \sum_{i\neq0}\, \hat{\mathbf{S}}_i \!\cdot \hat{\mathbf{s}}_i 
    + U\!\!\!\!\sum_{i\neq0,\sigma,\sigma^\prime} \!\!\!\!\hat{N}_{i\sigma}\hat{n}_{i\sigma^\prime}\,,
    \end{split}
\end{equation}
where $\smash{\hat{d}^\dagger_{i\sigma} (\hat{d}^{\vphantom{\dagger}}_{j\sigma})}$ creates (annihilates) an $|e\rangle$-atom on lattice site~$i$ with spin $\sigma$, and $\smash{\hat{N}_{i\sigma}= \hat{d}^\dagger_{i\sigma} \hat{d}^{\vphantom{\dagger}}_{i\sigma}}$ is the $e$-number operator on site $i$.
Since the impurity is delocalized over the lattice, the $U$ term can not anymore be compensated by adjusting the chemical potential on a specific site.
We thus study the system with $U=-2.75\,J$ for different values of $\tilde{t} \ll t$.
While for $\tilde{t}=0$ such large and positive~$U$ hinders the KE [see Fig.~\ref{fig1}(b)] , we find that for $\tilde{t} > 0$ a low-temperature Kondo scale is restored.
In particular, setting $\tilde{t} > 0$ leads to effects reminiscent of the Kondo lattice even with a single impurity, exhibiting a very large effective mass~$m^*\propto \chi_{zz}(T\to0)$, i.e., heavy-fermion behavior~\cite{Coleman2007,Hewson1993,Foss-Feig2010}.  
As shown in Fig.~\ref{fig4} for the specific heat, a novel resonance---completely absent for $\tilde{t}=0$---appears at low temperatures increasing monotonically with $\tilde{t}$.

\section{Conclusions and perspectives}\label{SecConclusions}
We have proposed an atom-array implementation of the Kondo impurity problem that leverages the ability to access the large-$J$ regime, where the Kondo cloud exhibits a quasi-local nature. We have numerically identified a favorable model parameter regime to stabilize the KE, as seen in various physical signatures that could be experimentally explored.
Our study will guide the realization of the KE in few-sites 1D systems within next-generation atomic quantum simulators, and could be readily generalized to two-dimensional systems.
While precise control of site- and state-selective optical potentials grants access to the pristine Kondo problem of a pinned magnetic impurity, we uncover novel low-temperature many-body effects when the impurity mobility is significant, deserving further investigations. %
Other future extensions of this work could address scenarios featuring multiple impurities fostering non-local magnetic exchanges~\cite{Jones1987,Affleck1992,Roch2008,Bayat2012,Pruser2014} and/or an interacting bath \cite{Furusaki1994,Wei2025}, as well as quantifying the entanglement properties in the Kondo cloud~\cite{Laflorencie2016,Bayat2012,Yoo2018,Bellomia2024PR}.
Optical tweezers may be utilized to pin multiple impurities in the system, suppressing their tunneling and the ensuing inelastic collisions, thus offering a promising route to the observation of fermion-mediated interactions.

%

\begin{acknowledgments}
We thank Jeff Maki and Oded Zilberberg for insightful discussions, and Pietro Massignan and Matteo Zaccanti for careful reading of this manuscript. A.A.~is indebted with Michele Fabrizio for many useful discussions and suggestions.
A.A. and M.C. acknowledge financial support from the National Recovery and Resilience Plan PNRR
MUR Project No.~PE0000023-NQSTI. 
A.R. acknowledges support by the Spanish Ministerio de Ciencia e Innovación (MCIN/AEI/10.13039/501100011033, grant PID2023-147469NB-C21), by the Generalitat de Catalunya (grant 2021 SGR 01411), and by the ICREA \textit{Academia} program. 
M.C. further acknowledges financial support from the National Recovery and Resilience Plan PNRR
MUR Project No.~CN00000013-ICSC and by MUR via PRIN 2020 (Prot.~2020JLZ52N-002) and PRIN 2022 (Prot.~20228YCYY7) programmes. 
F.S. acknowledges funding from the European Research Council (ERC) under the European Union’s Horizon 2020 research and innovation programme (project OrbiDynaMIQs, GA No.~949438), and from the Italian MUR under the FARE 2020 programme (project FastOrbit, Prot.~R20WNHFNKF) and the PRIN 2022 programme (project CoQuS, Prot.~2022ATM8FY).
\end{acknowledgments}


\appendix

\section{Derivation of the atomic Kondo model}\label{Sec:DerivModel}
The starting point to derive the atomic Kondo model studied in this work is the two-orbital Fermi-Hubbard model in one dimension describing the low-energy properties of cold AEAs in a state-dependent optical lattice~\cite{Gorshkov2010}:
\begin{widetext}
\begin{align}\label{eq:ham-two-orb}
\begin{split}
    \hat{\mathcal{H}} &=
        - t\sum_{\mathclap{i, \sigma\in\{\downarrow,\uparrow\}}}\left(\hat{c}^\dagger_{i \sigma} \hat{c}_{(i+1)\sigma} +
          \text{h.c.} \right)
        - \tilde{t}\sum_{\mathclap{i, \sigma\in\{\downarrow,\uparrow\}}} \left(\hat{d}^\dagger_{i \sigma} \hat{d}_{(i+1)\sigma} + \text{h.c.} \right)
        + U_\text{bath}\!\!\sum_{i,\sigma\neq\sigma^\prime} \hat{n}_{i\sigma} \hat{n}_{i\sigma^\prime} \\
        &\quad+ V_\mathrm{dir} \sum_{i, \sigma, \sigma^\prime} \hat{n}_{i\sigma} \hat{N}_{i\sigma^\prime}
        + V_\mathrm{ex} \sum_{i, \sigma, \sigma^\prime} \hat{c}_{i\sigma}^\dagger \hat{d}_{i\sigma^\prime}^\dagger \hat{c}_{i\sigma^\prime} \hat{d}_{i\sigma} 
        - \frac{B_z}{2}\sum_{i}\left(\delta \hat{m}_i + \Delta \hat{M}_i\right)
        + \mu\hat{n}_{0} + \epsilon\hat{N}_{0}.
\end{split}
\end{align}    
\end{widetext}

\noindent Here, $\hat{c}^\dagger_{i\sigma}$, $\hat{d}^\dagger_{i\sigma}$
($\hat{c}_{i\sigma}$, $\hat{d}_{i\sigma}$) are the fermionic creation
(annihilation) operators an atom at site $i$ in, respectively, the {\it
  orbital} state $\ket{g}$ or $\ket{e}$, and in the spin state $\sigma=\{\downarrow,\uparrow\}$.
The corresponding number operators are, respectively,
$\hat{n}_{i\sigma} = \hat{c}^\dagger_{i\sigma}\hat{c}_{i\sigma}$ and
$\hat{N}_{i\sigma} = \hat{d}^\dagger_{i\sigma}\hat{d}_{i\sigma}$.
$\hat{m}_i=\hat{n}_{i\uparrow}-\hat{n}_{i\downarrow}$ and
$\hat{M}_i=\hat{N}_{i\uparrow}-\hat{N}_{i\downarrow}$ are the spin
magnetization operators for the $\ket{g}$ and $\ket{e}$ states,
respectively. 
We consider half-filling occupation of the $\ket{g}$ states 
and only a single impurity in the $\ket{e}$ state, i.e. $ \sum_{i\sigma} \left\langle N_{i\sigma}
\right\rangle = 1$.
The local interaction strengths depend on the intra-orbital scattering
length $a_{gg}$ and the two inter-orbital scattering lengths $a_{eg}^\pm$, corresponding to the two-atom states
$\ket{gg}$ and $\ket{eg^\pm} = \frac{1}{2} \left(\ket{ge} \pm \ket{eg} \right) \otimes \left( \ket{\uparrow\downarrow} \mp \ket{\downarrow\uparrow} \right)$, respectively, and the onsite wavefunction overlap~\cite{Gorshkov2010,Scazza2014}. They can be expressed as $U_\text{bath} = {4\pi\hbar^2 a_{gg}}/{M} \int d\mathbf{r}\,{|\psi_g(\mathbf{r})|}^{4}$ and $U_{eg}^\pm = 4\pi\hbar^2 a_{eg}^\pm /M \int d\mathbf{r}\,{|\psi_g(\mathbf{r})|}^{2} {|\psi_e(\mathbf{r})|}^{2}$ --- an approximation justified for typical external confinements in experiments~\cite{Bloch2008}. Here, $M$ is the atomic mass, $\psi_{g,e}(\mathbf{r})$ denotes the on-site spatial wavefunction of a $g,e$ atom as determined by the state-dependent optical confinement in the experiment. 
Specifically, one obtains $V_{\rm dir}={{1\over2} (U^+_{eg}+U^-_{eg})}$ (direct term) and $V_{\rm ex}={{1\over2}(U^+_{eg}-U^-_{eg})}$ (exchange term)~\cite{Gorshkov2010,Scazza2014}.
It was found for ${}^{171}$Yb that $U^\pm_{eg}>0$ ~\cite{Ono2019,Bettermann2020}, and the following relation holds: $0 < |V_{\rm ex}| <
V_{\rm dir}$.
The other model parameters~$t$ and $\tilde{t} \ll t$ can be determined in a tight-binding band-structure calculation~\cite{Bloch2008}. 
The last four contributions in Eq.~\eqref{eq:ham-two-orb} represent optional terms,
i.e., orbital-dependent Zeeman shifts $\delta/2$ and $\Delta/2$ in the
presence of a finite magnetic field $B_z \neq 0$ as well as 
orbital-dependent energy potentials $\mu$ and $\epsilon$ acting on atoms at a specific lattice site labeled~$i=0$ (i.e., the impurity site).

In order to derive a Kondo-like Hamiltonian, we examine the first terms in the second line of Eq.~\eqref{eq:ham-two-orb}, which describe the inter-orbital interactions,
\begin{align}\label{eq:ham-interorb}
    \hat{\mathcal{H}}_\text{orb} = 
        V_\mathrm{dir} \sum_{i, \sigma, \sigma^\prime} \hat{n}_{i\sigma} \hat{N}_{i\sigma^\prime}
        + V_\mathrm{ex} \sum_{i, \sigma, \sigma^\prime} \hat{c}_{i\sigma}^\dagger \hat{d}_{i\sigma^\prime}^\dagger \hat{c}_{i\sigma^\prime} \hat{d}_{i\sigma}.
\end{align}
Then, we consider the discrete Fierz identity
$\delta_{\sigma\gamma}\, \delta_{\gamma^\prime\! \sigma^\prime}=
(\delta_{\sigma\!\sigma^\prime} \delta_{\gamma^\prime \!\gamma} +
\boldsymbol{\tau}_{\sigma\!\sigma^\prime}\cdot \boldsymbol{\tau}_{\gamma^\prime \!
  \gamma})/2$ (where $\delta_{ij}$ is the Kronecker symbol) to rewrite the exchange interaction in terms
of Abrikosov's pseudospin representations
$\hat{\boldsymbol{s}}_{i} = {1 \over 2}
\sum_{\sigma\!,\sigma^\prime} \hat{c}^\dagger_{i\sigma}
\boldsymbol{\tau}_{\sigma\!\sigma^\prime} \hat{c}_{i\sigma^\prime}$ and
$\hat{\boldsymbol{S}}_{i}~=~{1 \over 2}
\sum_{\sigma\!,\sigma^\prime} \hat{d}^\dagger_{i\sigma}
\boldsymbol{\tau}_{\sigma \!\sigma^\prime} \hat{d}_{i\sigma^\prime}$ (where $\boldsymbol{\tau} = \{\sigma_x, \sigma_y, \sigma_z\}$ are the Pauli matrices) as:
\begin{widetext}
\begin{align}
\begin{split}
    \sum_{\sigma,\sigma^\prime} \hat{c}^\dagger_{\sigma} \hat{d}^\dagger_{\sigma^\prime} \hat{c}_{\sigma^\prime} \hat{d}_{\sigma}
    &= \sum_{\sigma,\sigma^\prime,\gamma,\gamma^\prime} \hat{c}^\dagger_{\sigma} \hat{d}_{\gamma} d_{\gamma^\prime}^\dagger c_{\sigma^\prime} \left( \delta_{\sigma\gamma} \delta_{\gamma^\prime \sigma^\prime} \right)\\
    &= \frac{1}{2} \sum_{\sigma,\sigma^\prime,\gamma,\gamma^\prime} \hat{c}^\dagger_{\sigma} \hat{d}_{\gamma} d_{\gamma^\prime}^\dagger c_{\sigma^\prime} \left( \delta_{\sigma\!\sigma^\prime} \delta_{\gamma^\prime\gamma} + \boldsymbol{\tau}_{\sigma\!\sigma^\prime} \cdot \boldsymbol{\tau}_{\gamma^\prime\gamma}\right) \\
    &= \frac{1}{2}\sum_{\sigma,\gamma} \hat{c}^\dagger_{\sigma} \hat{d}_{\gamma} d_{\gamma}^\dagger c_{\sigma}
        + \frac{1}{2} \sum_{\sigma,\sigma^\prime,\gamma,\gamma^\prime} \left(\hat{c}^\dagger_{\sigma} \boldsymbol{\tau}_{\sigma\!\sigma^\prime} \hat{c}_{\sigma^\prime} \right) \cdot \left(\hat{d}_{\gamma} \boldsymbol{\tau}_{\gamma^\prime\gamma} \hat{d}^\dagger_{\gamma^\prime}\right) \\
    &= \frac{1}{2}\sum_{\sigma,\gamma} \hat{n}_{\sigma} \left(1 - \hat{N}_{\gamma} \right)
        - \sum_{\sigma,\sigma^\prime,\gamma,\gamma^\prime}
            \left(\hat{c}^\dagger_{\sigma} \boldsymbol{\tau}_{\sigma\!\sigma^\prime} \hat{c}_{\sigma^\prime}\right)
            \cdot \left(\frac{1}{2}\hat{d}^\dagger_{\gamma^\prime} \boldsymbol{\tau}_{\gamma^\prime\gamma} \hat{d}_{\gamma}\right) \\
    &= \sum_\sigma \hat{n}_{\sigma}
        - \frac{1}{2} \sum_{\sigma, \sigma^\prime} \hat{n}_{\sigma} \hat{N}_{\sigma^\prime}
        - 2 \hat{\boldsymbol{s}} \cdot \hat{\boldsymbol{S}}.
\end{split}
\end{align}
\end{widetext}
Note that to ease notation, we dropped the site index~$i$ due to the local nature of Eq.~\eqref{eq:ham-interorb}.
Using this result, we can now rewrite the inter-orbital interaction Hamiltonian $\mathcal{H}_\text{orb}$ as:
\begin{widetext}
\begin{align}
\begin{split}
    \hat{\mathcal{H}}_\text{orb} &= 
        V_\mathrm{dir} \sum_{i, \sigma, \sigma^\prime} \hat{n}_{i\sigma} \hat{N}_{i\sigma^\prime}
        + V_\mathrm{ex}
            \sum_{i,\sigma} \hat{n}_{\sigma}
            - \frac{V_\mathrm{ex}}{2} \sum_{i,\sigma,\sigma^\prime} \hat{n}_{i\sigma} \hat{N}_{i\sigma^\prime}
            - 2 V_\mathrm{ex} \sum_{i} \hat{\boldsymbol{s}}_{i} \cdot \hat{\boldsymbol{S}}_{i}\\
        &= \left(V_\mathrm{dir} - \frac{J}{4}\right) \sum_{i, \sigma, \sigma^\prime} \hat{n}_{i\sigma} \hat{N}_{i\sigma^\prime}
            - J \sum_{i} \hat{\boldsymbol{s}}_{i} \cdot \hat{\boldsymbol{S}}_{i}
            + \frac{J}{2} \sum_{i,\sigma} \hat{n}_{i\sigma}.
\end{split}
\end{align}
\end{widetext}
in which we made use of the implicit relation $J = 2V_\mathrm{ex}$ between the Kondo spin-exchange coupling $J$ and the atomic inter-orbital spin-exchange energy $V_\mathrm{ex}$.
Omitting the irrelevant linear term $(J/2)\sum_{i,\sigma} \hat{n}_{i\sigma}$, we finally obtain the following Kondo-like Hamiltonian for the atomic system:
\begin{align}\label{Hh}
    \hat{\mathcal{H}} =
        \hat{\mathcal{H}}_K + \hat{\mathcal{H}}^\prime  - B_z\sum_{i}\left(\delta \hat{\boldsymbol{s}}_{zi} + \Delta \hat{\boldsymbol{S}}_{zi}\right)
        + \mu\hat{n}_{0} + \epsilon\hat{N}_{0}.
\end{align}
The first term represents the  well-known textbook Kondo Hamiltonian~\cite{Hewson1993,Coleman2007}
\begin{align}\label{Hk}
    \hat{\mathcal{H}}_K =
        - t \sum_{\mathclap{i,\sigma,\sigma^\prime}} \left( \hat{c}_{i\sigma}^\dagger \hat{c}_{(i+1) \sigma} + \text{h.c.} \right)
        - J \sum_{i} \hat{\boldsymbol{s}}_{i} \cdot \hat{\boldsymbol{S}}_{i},
\end{align}
while  the term $\hat{\mathcal{H}}^\prime$ describes deviations of the atomic system from a standard Kondo model %
\begin{equation}\label{dHk}    
\begin{split}
    \hat{\mathcal{H}}^\prime &= 
        - \tilde{t} \sum_{\mathclap{i,\sigma,\sigma^\prime}} \left( \hat{d}_{i\sigma}^\dagger \hat{d}_{(i+1) \sigma} + \text{h.c.} \right)\\
        &\phantom{=}+ U_\text{bath} \sum_{\mathclap{i,\sigma\neq\sigma^\prime}} \hat{n}_{i\sigma} \hat{n}_{i\sigma^\prime}
        + \left(V_\mathrm{dir} - \frac{J}{4}\right) \sum_{i, \sigma, \sigma^\prime} \hat{n}_{i\sigma} \hat{N}_{i\sigma^\prime}.
\end{split}
\end{equation}

The model Eq.~(\ref{Hh}) can be further reduced to obtain the effective
single-impurity Kondo Hamiltonians [Eqs.~(1)-(2)] discussed in the main
text.
Motivated by its vanishingly small contribution for ${}^{171}$Yb~\cite{Kitagawa2008}, we begin by neglecting the intra-orbital $\ket{g}$-state interaction term of strength $U_\text{bath}\ll t$. In one dimension the inclusion of this interaction term dramatically changes the nature of the metallic state giving rise to Luttinger-Kondo effect~\cite{Furusaki1994,Furusaki2005}, 
whose investigation is beyond the scope of this work.  

Next, we analyze the inter-orbital interaction term in
Eq.~(\ref{dHk}). Recalling the defining relation for the Kondo coupling $J=2V_{\rm ex}$ and the definitions of $V_{\rm dir}$ and $V_{\rm ex}$, we can recast
the coupling in the form: 
$$
V_{\rm dir} - \tfrac{1}{4}J = \tfrac{1}{4}U^+_{eg} +
\tfrac{3}{4}U^-_{eg} \defeq U\,.
$$
In the last relation we implicitly
defined the interaction strength $U$ [see Eq.~(1) in the main text)]. The
previous relation clarifies the origin of this coupling term,
and makes it transparent that this quantity directly depends on the atomic scattering lengths. The exact values of both $U$ and $J$ also depends on the external confinement of the $\ket{g}$ and $\ket{e}$ atoms. In particular, $U$ and $J$ can be tuned independently of other parameters by acting on the optical potentials providing confinement along the directions orthogonal to the 1D array. Yet, the ratio $U/|J|\approx 2.75$ for the case of ${}^\text{171}\text{Yb}$ as discussed in the main text is fixed by the value of the experimentally measured scattering lengths $a_{eg}^+ \simeq 240\, a_0$ and $a_{eg}^- \simeq 389\, a_0$~\cite{Bettermann2020} ($a_0$ is the Bohr radius), irrespective of confining potentials.

Finally, we discuss how to tune the local potentials
$\mu n_{0} + \epsilon N_{0}$ to compensate (completely or
partially) for the effects of the two remaining perturbations, namely the inter-orbital repulsion term  $\propto U$ and the $\ket{e}$-state hopping term~$\propto \tilde{t}$.
This transforms the model in Eq.~(\ref{Hh}) into the effective Kondo problem in Eq.~(1).
First, we note that for $\tilde{t}=0$ the fermion populating the
impurity occupies the $\ket{e}$-state at an arbitrary site. In the absence
of any charge fluctuation, the inter-orbital repulsion term
$\propto U$ effectively reduces to a quadratic term proportional to the occupation
of the conduction band states $n_{i}$. Any such term can be
reabsorbed with a redefinition of the local potential $\mu$. 
Thus, the problem of deriving an effective Kondo model as in Eq.~(1) reduces to tuning the
remaining potential $\epsilon$ so to localize the impurity fermion at a
given site $i=0$. As the motion of the impurity fermion is affected by the
Kondo coupling to the conduction-band fermions, the energy gain required
for the localization should be compared not just to the bare bandwidth of the $\ket{e}$-states ($4\tilde{t}$) but also to that of the conduction fermions
($4t$). Choosing $\epsilon \geq 4(\tilde{t}+t)$ ensures exponential
localization of the impurity fermion at the site $i=0$ with vanishing
charge fluctuations up to the largest accessible temperature.
By tuning the local potential $\mu$ we effectively adjust the value of $U$. 
The data in Fig.~1 and Fig.~2 of the main text have been obtained using this setup. In particular, the results shown in Fig.~2(b,c) demonstrate a perfect match between the compensated model Eq.~(\ref{Hh}) and the plain vanilla Kondo problem solved independently. 

In the main text, we have also discussed Eq.~(1) in the context of the proposed implementation where the optical lattice is seen solely by the $\ket{g}$-states, and the impurity experiences just a deep local confining potential $\epsilon$ at the impurity site. In this particular case, the tunnelling strength $\tilde{t}$ of the impurity is strictly zero, and Eq.~(1) becomes an exact description of the experimental scenario with $U$ being an effective tuning parameter through the adjustment of the $\ket{g}$-state potential $\mu$ at the impurity site.

\section{Numerical methods} \label{Sec:NumericalMethods}
\subsection{Exact diagonalization}
We solve the two main models presented in the main text [see Eq.~(1) and Eq.~(2)] using a Lanczos-based exact diagonalization method~\cite{Lin1993,Amaricci2022,CrippaSciPost2025}. The numerical approach developed for this work relies on a suitable extension of the massively parallel algorithm of Ref.~[\onlinecite{Amaricci2022}] and available online \footnote{An open-source version of the implemented numerical method can be found at \url{https://github.com/EDIpack/EDIkondo}}.
This method allows to compute a large number of eigen-solutions of the quantum many-body Hamiltonian in the low-energy part of the spectrum, notwithstanding the exponentially large dimension of the Fock space.  
In addition, the use of finite-temperature Krylov-subspace algorithms allows for the evaluation of dynamical response functions within the framework of linear response theory.

We consider a 1-dimensional lattice with $L$ sites, each hosting two levels, $\ket{g}$ and $\ket{e}$, with periodic boundary conditions. The local Hilbert space $\mathcal{H}$ at each lattice site is given by the tensor product $\mathcal{H}_g\otimes \mathcal{H}_e$, where $\mathcal{H}_g=\{\ket{0},\ket{\up},\ket{\dw},\ket{\up\dw} \}$ is the fermionic Hilbert space of the $\ket{g}$ states and $\mathcal{H}_e=\{\ket{\up},\ket{\dw}\}$ is that of the $\ket{e}$ impurities.
The total Fock space is given by $\mathcal{F}=\bigoplus_{n=0}^{N} S_- \mathcal{H}^{\otimes n}$, where $S_-$ is the anti-symmetrization operator and $N=N_g+N_e$ is the total number of fermions. In this work, we focus on the case $N_e=1$ and $N_g=L$, though this discussion applies as well to the presence of multiple impurity fermions.
The full many-body Hilbert space has dimension $\mathcal{N}=2^{2N_g} \times 2^{N_e}$, which obviously becomes intractable already for few sites. To reduce the computational complexity of the problem, we exploit the conservation of spin-resolved occupation numbers labeling sectors by the quantum numbers $\vec{Q}=\{N^\up_g,N^\dw_g,N_e \}$.
The presence of these symmetries allows one to decompose the Hamiltonian into separated blocks, each with dimensions $D_{\vec{Q}}=\sum_{n_e=0}^{N_e=1} \binom{L}{n_e}\binom{L}{N^\up_g-n_e}\binom{L}{N^\dw_g-(N_e-n_e)} \ll \mathcal{N}$ which can be diagonalized independently.
We employ a refined Arpack-Lanczos diagonalization algorithm~\cite{Lehoucq1998,Amaricci2022} to obtain the $M_{\vec{Q}}$ lowest eigen-pairs in each sector, collecting enough states to fulfill the condition $e^{-E_M/k_BT}<\varepsilon$, where $M=\sum_{\vec{Q}}M_{\vec{Q}}$ and $\varepsilon$ is a sufficiently small energy cutoff.
The obtained list of eigenstates $\{\ket{\psi_m}\}_{m=0,\dots M}$ can then be used to evaluate any observable $\langle \hat{\mathcal{O}}\rangle =\sum_{m=1}^{M}e^{-E_m/k_BT}\bra{\psi_m}\hat{\mathcal{O}} \ket{\psi_m}$, including the internal and the kinetic energy of the system. Analogously, using the Krylov method, it is possible to evaluate any dynamical correlation function of the form $\langle \hat{A}(t-t') \hat{B}(0)\rangle=\sum_{m=1}^{M} e^{-E_m/k_BT}\bra{\psi_m}\hat{A}(t-t') \hat{B}(0) \ket{\psi_m}$.

\begin{figure*}
\includegraphics[width=\linewidth]{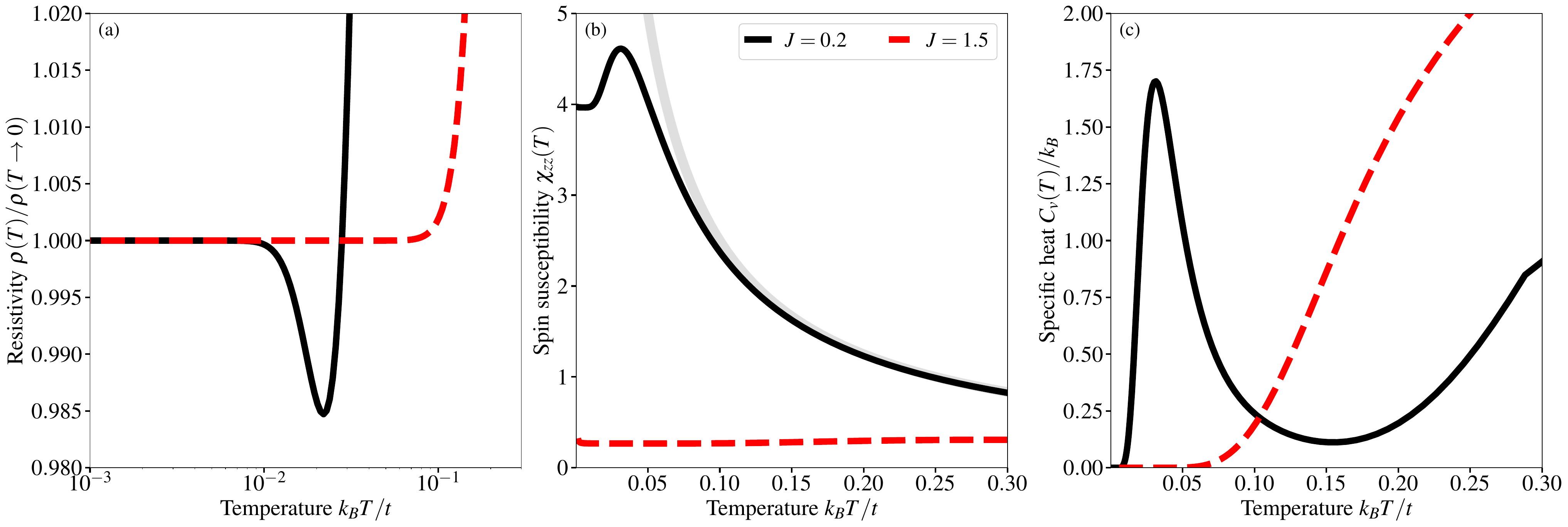}
\caption{\label{figSM1}%
  {\bf The effects of weak and strong coupling $J$ in the Kondo problem}.
  (a) Temperature behavior of the normalized resistivity
  $\rho(T)/\rho(T\to 0)$, showing the formation of a marked minimum
  due to the Kondo effect for $|J|=0.2\,t$. 
  (b) Local spin susceptibility $\chi_{zz}(T)$ as a function of the
  temperature $T$. Results for $|J|=0.2\,t$ display a crossover from a Curie-Weiss
  $\tfrac{1}{4k_BT}$ (gray solid line)  to a Pauli-like constant
  behavior at low temperature.
  (c) Temperature-dependence of the specific heat $C_v(T)$.
  All data are for $|J|=0.2\,t$ (black solid) and $|J|=1.5\,t$ (red dashed).}
\end{figure*}

\section{Evaluating the different Kondo markers}

The Kondo effect is accompanied by different remarkable changes in the behavior of the system which impacts transport, spin or thermodynamic properties. In the following sections we discuss in more details the evaluation of different markers which can qualitatively assess the onset of the Kondo screening through a crossover from high to low temperature in a few-body lattice system.  

\subsection{Resistivity}
In the case of diluted impurities in solids, the Kondo effect is conventionally related to the existence of a minimum in the resistivity $\rho(T)$~\cite{Hewson1993}. The presence of such minimum is due to the logarithmic increase of the metallic resistivity at low temperatures, in turn caused by the scattering processes involving the exchange of spin states between the impurity and the mobile electrons. 
While genuine logarithmic corrections can only arise in the thermodynamic limit, we define, for our finite system, a Kondo scale $T_K^{(\rho)}$ as the temperature at which the electrical resistivity attains its minimum value: $\min_T \rho(T) = \rho(T_K^{(\rho)})$.

To characterize the conduction properties of our system, we compute the Drude weight $D$, which quantifies the system's asymptotic response to a uniform electromagnetic field. The Drude weight is defined as the singular part of the optical conductivity $\sigma(\mathbf{q}=0,\omega)=D\delta(\omega)+\sigma_\mathrm{reg}(0,\omega)$~\cite{Fye1991,Giamarchi1995,Richaud2021}, which relates the current ${\bf J}$ to the vector potential ${\bf A}$ via ${\bf J}(\mathbf{q},\omega) = \sigma(\mathbf{q},\omega) {\bf A}(\mathbf{q},\omega)$.  
In metals~\cite{Fye1991} or integrable systems~\cite{Zotos2002}, the Drude weight is positive, indicating ballistic transport. In contrast, a vanishing Drude weight signals insulating behavior~\cite{Fye1991} or diffusive transport~\cite{Zotos2002}. 

The resistivity $\rho=\lim_{\omega\rightarrow0}1/\sigma(0,\omega)$ is thus obtained from the knowledge of the Drude weight which can only be accessed indirectly through linear-response theory~\cite{Eder1996PR}. 
We consider atoms on a lattice subject to a small perturbing electro-magnetic field. The latter enters the model through the Peierls' substitution $t \rightarrow te^{-i {\mathbf A}(t)\cdot (R_i-R_j)}$ where $R_{i,j}$ are lattice vectors for two nearest-neighbor sites. Expanding in powers of ${\bf A}$, the kinetic energy becomes $\hat{T} = \hat{T}_0 - \hat{{\mathbf j}}\cdot{\mathbf A} + \frac{\hat{T}_0 {\mathbf A}^2}{2} + \dots$
with the bare kinetic energy and current density operators defined as:
\begin{equation}
   \begin{split}
    &\hat{T}_0 = -t \sum_{\langle
      ij\rangle,\sigma}\left( \hat{c}^\dagger_{i\sigma}\hat{c}_{j\sigma} + \text{h.c.}\right),\\
    &\hat{{\mathbf j}} = -i \sum_{i,\sigma}\left(\hat{c}^\dagger_{i\sigma}\hat{c}_{i+{\mathbf e}\sigma}-\hat{c}^\dagger_{i+{\mathbf e}\sigma}\hat{c}_{i\sigma}\right),
   \end{split}
\end{equation}
where ${\bf e}$ is the unit vector along the direction of the current ${\bf j}$. The total current is obtained by differentiating $\hat{T}$ with respect to the vector potential~\cite{Fye1991}: $\hat{{\mathbf J}} = -\tfrac{\delta \hat{T}}{\delta{\mathbf A}(t)}$. Taking thermal expectation values, we obtain an expression for the total current: $\langle \hat{{\mathbf J}}(t)\rangle = \langle \hat{{\mathbf j}}(t)\rangle - \langle {\hat{T}_0}\rangle{\mathbf A}(t)$. 
Using linear-response theory and the Green-Kubo relation, we can express the conductivity in terms of a current correlation and the diamagnetic contribution which, after a Fourier transform to wave-vector and frequency domain, reads:
$$
\sigma(\mathbf{q},\omega) = \frac{1}{\omega+i\eta}\left( \chi(\mathbf{q},\omega) -i
  \langle \hat{T}_0 \rangle \right)
$$
where:
$$
\chi(\mathbf{q},\omega) = -i\lim_{\eta\to 0^+}\int_0^\infty dt\, e^{-i\omega t - i\eta t} \left\langle[ \hat{{\bf j}}(\mathbf{q},t), \hat{{\bf j}}(-\mathbf{q},0)] \right\rangle
$$
is the retarded current susceptibility. 
While in principle both terms in the previous relation can be efficiently evaluated with full vector and frequency dependence, for our purposes it is enough to estimate the Drude weight as 
\begin{equation}
  D = -\frac{\pi}{L}\left[ \langle \hat{T}_0
    \rangle - \chi(0,\omega\to0) \right]  
\end{equation}
where $L$ is the system size. This expression provides a relation between $D$ and the intrinsic dynamical response of the system. Using the Krylov method it is straightforward to construct the thermal expectation value of the current-current correlation function entering the expression of $D$:
\begin{widetext}
\begin{equation}
  \chi(0,\omega\to 0) =  \lim_{\omega\to 0}\frac{1}{Z} \sum_{m=1}^{M} e^{-E_m/k_B T}\langle{\psi_m}|\hat{\bf j}\left(\omega - (H-E_m) \right)^{-1} \hat{\bf j}|{\psi_m}\rangle
  =\lim_{\omega\to 0}\frac{1}{Z} \sum_{m=1}^{M} e^{-E_m/k_B T}\sum_{i=1}^{K_i}  \frac{| \langle \psi_i | \hat{\bf j} | \psi_m \rangle|^2}{E_m - E_i - \omega + i\eta}
\end{equation}
\end{widetext}
where $|\psi_m\rangle$ and $E_m$ are the eigenstates and eigenvalues of the Hamiltonian and the second sum is truncated to the number $K_i$ of the Krylov basis, i.e. the number of excitations with a sizable overlap with the starting state $\ket{\psi_m}$; $Z=\sum_{m=1}^{M} e^{-E_m/k_B T}$ is the truncated partition function.

\begin{figure*}
\includegraphics[width=\textwidth]{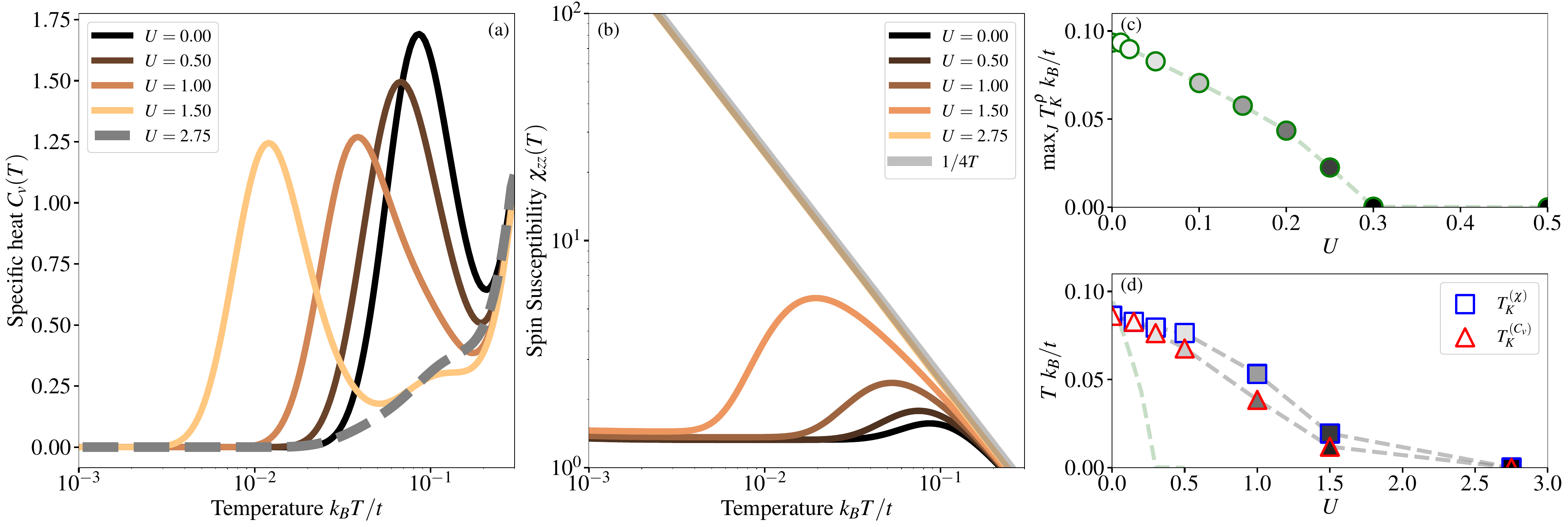}
\caption{\label{figSM2}%
  {\bf Effects of the inter-orbital repulsion $U>0$ to the
  Kondo problem}. Data are obtained for a chain of length $L=7$ and $|J|/t=0.50$.  
  Evolution of the (a) specific heat $C_v$ and (b) spin susceptibility
  $\chi_{zz}$ as a function of temperature. The measured scattering properties for ${}^{171}$Yb fix $U\simeq2.75\,|J|$
  which completely suppresses the Kondo effect. 
  Right panels show the reduction of the (c) \smash{maximum Kondo temperature
  ${\rm max}_J\; T_K^{(\rho)}$}, and (d) the specific heat (red symbol) and spin
  susceptibility (blue symbols) Kondo temperature estimate
  $\smash{T_K^{(C)}}$ and $\smash{T_K^{(\chi)}}$ for $|J|/t=0.50$. 
}
\end{figure*}

\subsection{Spin susceptibility}
A distinctive feature of the Kondo effect is the 
increasing interaction among the spins of the localized and mobile fermions at low temperatures. The dynamical build-up of the Kondo screening is captured by the temperature evolution of the local spin susceptibility $\chi_{zz}(T)$. 
This quantity displays a characteristic high-temperature Curie-like behavior, signaling the existence of free local moments. Yet, upon lowering the temperature, as the impurity spin gets progressively screened, the spin susceptibility reaches a maximum value and sets into a Pauli-like behavior with a strongly renormalized effective mass. The crossover between these two regimes sets the Kondo temperature scale. 

The local spin susceptibility is defined as:
\begin{equation}
    \chi_{zz}(T) = \int_0^{1/k_BT} d\tau \chi_{zz}(\tau)
\end{equation}
where, using linear-response theory, the spin-spin dynamical correlation function 
\begin{equation*}
\begin{split}
  \chi_{zz}(\tau) & = \langle [\hat{S}_{z0}(\tau), \hat{S}_{z0}(0)]
  \rangle\\
  & =  \frac{1}{Z} \sum_{m=1}^{M} \sum_{n=1}^{K_i}  
  \!(1\!-\!e^{-\Delta E_{mn}/k_BT})e^{-\Delta E_{mn}\tau}
  |\langle \psi_n|\hat{S}_{z0}|\psi_m\rangle|^2
  \end{split}   
\end{equation*}
is evaluated using the Krylov sub-space method. $\hat{S}_{z0}$ is the $z$ component for the impurity spin operator, $\ket{\psi_{i}}$ are the eigenstates of the Hamiltonian with energies $E_i$, $\Delta E_{mn}=E_m - E_n$, and $K_i$ is the Krylov basis dimension. 

An example illustrating the different Kondo-physics regimes in terms of the temperature
evolution of the spin susceptibility is reported in \figu{figSM1}. The smaller investigated value of the coupling $J$ features a Curie-Weiss increase at high temperature eventually saturating to a Pauli-like behavior at low temperatures with a large effective mass set by the residual value at zero temperature.
In contrast, the largest value of $J$ features, for all temperatures, an essentially constant Pauli behaviour with a small effective mass. 

\subsection{Specific heat and entropy} 
The restoration of a local Fermi liquid upon reducing the temperature below the Kondo scale gives rise to a large anomalous increase of the specific heat $C_v$, which is associated to the quenching of the many-body atomic states contributing to the formation of the narrow Kondo resonance. The presence of such low-temperature {\it Schottky anomaly} in the specific heat provides a signature of the Kondo physics and an indirect estimate of the associated Kondo temperature.
The specific heat is defined as: 
\begin{equation}
  C_v = -\frac{1}{k_B}\frac{\mathrm{d} \langle \hat{H} \rangle}{\mathrm{d}T},
\end{equation}
where $\hat{H}$ is the model Hamiltonian operator, including kinetic and interaction terms and $\langle{\hat{H}}\rangle$ is the internal energy of the system.

The fluctuation-dissipation theorem relates the behavior of the specific heat to the entropy $S$: 
\begin{equation}
  S(T) = S_0 + \int_0^T \frac{C_v(T^\prime)}{T^\prime} dT^\prime
\end{equation}
where $S_0$ is a reference entropy, often set to zero at the lowest computed temperature. Fermi-liquid theory dictates a linear vanishing behavior of the entropy at low temperature, i.e. in the fully-screened-impurity regime. The crossover from this regime to that of the unscreened moments is associated to the entropy reaching the value $S=\ln{2}$, reflecting the Kondo quenching of the impurity spins associated to the Schottky anomaly peak. 

In \figu{figSM1}(c) we report the behavior of the specific heat $C_v(T)$ and the corresponding entropy curve $S(T)$ for two values of the Kondo coupling $J$ in and away from the Kondo regime. The smaller value of $J$ features the presence of a well defined anomaly at low temperatures, which is totally absent for larger~$J$.

\begin{figure*}
\includegraphics[width=0.64\textwidth]{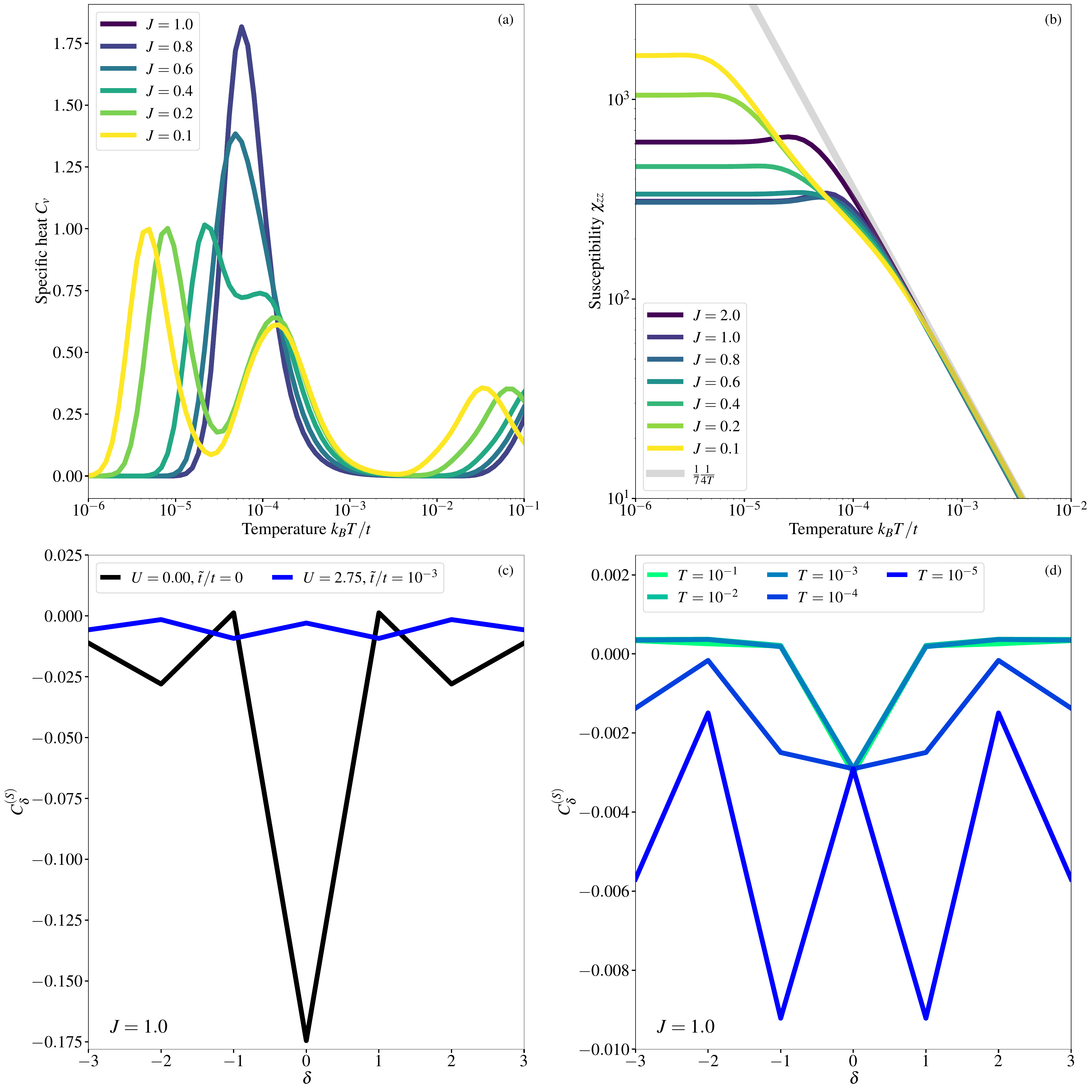}
\caption{
  {\bf Emergent Kondo lattice physics from a single tunneling impurity}.
  (a-b) Evolution of the specific heat $C_v$ (a) and impurity spin susceptibility $\chi_{zz}$ (b) as a function of
  temperature and Kondo coupling $J$.
  (c) Spatial profile of spin-spin correlations
 \smash{$\mathcal{C}^{(S)}_\delta$} for
  $|J|=1.0$ and $T=0$. The figure compares the single Kondo (black solid line) with the Kondo
  lattice (blue solid line) behavior.
  (d) \smash{$\mathcal{C}^{(S)}_\delta$} profile for $|J|=1.0$ as a function of 
  temperature. 
  All data for the Kondo lattice case are for a chain of length
  $L=7$, $\tilde{t}=10^{-3}\,t$ and $U=2.75\,|J|$.
}
\label{figSM4}%
\end{figure*}

\section{Finite intra-orbital density-density repulsion}\label{subSecUgt0}
We discuss here, in more detail, the impact of the inter-orbital interaction $U$ on the Kondo effect in the single-impurity problem. We tune the value of $U$  between the two regimes $U/t=0$, corresponding to a complete compensation, and $U/|J|=2.75$ corresponding to the absence of compensation.

In \figu{fig1}(b) we presented the evolution with $U$ of the Kondo temperature $T_K^{(\rho)}$ as estimated from the resistivity minimum. Here, to better appreciate the suppression of the Kondo effect driven by $U$, we complement the analysis considering other markers other than the charge transport. The results are reported in \figu{figSM2}.
In panel \figu{figSM2}(a) we show the specific heat as a function of temperature for different values of $U$ and for $J/t=0.5$, see \figu{fig1}(b)-(c).
The clearly visible Schottky anomaly peak gradually decreases in height while shifting to lower temperatures as $U$ increases. 
Remarkably, values of the repulsion strength $U$ around the ${}^{171}$Yb estimated value of $U/|J|=2.75$ lead to a complete suppression of the peak and, thus, of the KE.  

The analysis of the spin susceptibility $\chi_{zz}$ further emphasizes this behavior, as illustrated in panel \figu{figSM2}(b). For zero and small values of $U$, the susceptibility exhibits the expected saturation, after a peak value, to a renormalized Pauli behavior. This trend indicates the progressive screening of the impurity local moment till a local Fermi-liquid state is reached. 
As $U$ increases, the peak of $\chi_{zz}$ shifts to lower temperatures, while the renormalized Pauli behavior remains almost unchanged. For values of $U\simeq2.75\,|J|$, we observe a complete suppression of the Kondo effect, with $\chi_{zz}$ following the ideal Curie law, $\chi_{zz}=\tfrac{1}{4T}$ characteristic of a free local moment.

To quantify the suppression of the Kondo effect driven by $U$, we report in panels \figu{figSM2}(c)-(d) the evolution of the Kondo temperature estimated from two complementary sets of observables.
Specifically, panel (c) shows the reduction of the maximum of the Kondo temperature as estimated from the resistivity minimum, see also \figu{fig1}. The reported values correspond to different Kondo couplings $J/t$. Our results show that, if one focuses on the current response, the KE completely disappears already for $U/|J|=0.30$.
However, the specific heat and the spin susceptibility estimates for $T_K$ reported in panel (d) display a different behavior.
At $U = 0$, different indicators yield nearly identical estimates of the Kondo temperature. At finite $U$, however, the estimated value becomes strongly dependent on the specific indicator employed [see panel (d) of \figu{figSM2}]. This discrepancy likely reflects the enhanced sensitivity of transport properties in one-dimensional systems, which leads to a partial decoupling among spin, transport, and thermodynamic responses. As a result, spin-related signatures of Kondo physics appear more robust than those derived from transport observables. In conclusion, the Kondo effect, understood as the quasi-local screening of the impurity’s local moment, persists up to relatively large interaction strengths $U\lesssim |J|$. However, as $U$ approaches the value of $2.75\,|J|$ estimated for $^{171}$Yb atoms, all indicators, regardless of their nature, consistently signal the complete suppression of Kondo physics.

\section{Emergent Kondo lattice behavior from a single mobile impurity}
Finally, we present additional results concerning the emergent
Kondo-lattice behavior observed in the presence of a tunneling $\ket{e}$-state impurity. 
As we discussed in the main text, the presence of a delocalized
impurity renders the compensation of the inter-orbital interaction term $U$ impossible, due to occupation fluctuations in both $\ket{e}$ and $\ket{g}$ states which cannot be compensated by adjusting the local potentials.
However, the term $U$ by suppressing charge fluctuations in the
$\ket{g}$ states ultimately lead to a partial restoration of the Kondo effect
which can be modulated through the tunneling amplitude $\tilde{t}$ (see \figu{fig4} in the main text).
Here we consider the case $\tilde{t}=10^{-3}\,t$ and $U=2.75\,|J|$. 
In this regime the resistivity $\rho$ does not exhibit any resistivity minimum, indicating that, in one dimension, the current-related
signature of the Kondo effect is lost.
Yet, the evolution of the specific heat and spin susceptibility shown in \figu{figSM4} confirms the emergence of a Kondo-lattice
behavior in this system.
For instance, in panel (a) we show various specific-heat curves as a function
of the temperature for increasing values of the Kondo coupling
$J$. Our results unequivocally show the formation of a robust Kondo peak
already for moderate values of $J$. Interestingly, the Schottky 
anomaly is split into two peaks in the weak-coupling regime as a consequence of
the large value of $U$ with respect to the Kondo scale~$J$. 

The analysis of the spin susceptibility
further consolidates this picture. The results reported in panel (b)
indicate a general crossover of the susceptibility curves from a
Curie-Weiss-like behavior to a Pauli-like behavior at extremely low temperatures. 
In the weak-coupling regime, such crossover takes place in two steps,
in agreement with the double-peak structure observed for $C_v$. First, the $\chi_{zz}$ curves bend away from the $\frac{1}{L}\frac{1}{4T}$ (corresponding to fractional free local moment),
signaling the incipient start of the screening process. However, only
at lower temperatures the susceptibility clearly signals the formation of a metallic state, in turn characterized by an extremely
large effective mass $m^*\propto \chi_{zz}(T\to0)$, i.e. heavy fermions behavior~\cite{Coleman2007,Hewson1993}.
In the intermediate-to-large coupling regime the susceptibility takes
on a more conventional Kondo behavior, showing a single-step crossover
and strongly renormalized Pauli behavior. 

Having clarified the emergence of a lattice Kondo effect in the one dimensional
chain, we now investigate the characteristics of the magnetic
correlations $C^{(S)}_\delta$ introduced in the main text. We consider the
spin-spin correlations with respect to the center (reference) site $\delta=0$,
although the results are easily extended to the any other sites in PBC.
To begin with, in \figu{figSM4}(c) we compare the $C^{(S)}_\delta$ profile
between the single and the lattice Kondo problems. The single impurity shows a characteristic symmetrically oscillating
and decaying behavior, i.e. Friedel oscillations, identified with the
formation of a Kondo cloud. Remarkably, in the lattice case the
profile of the spin correlations between $\ket{e}$ impurity and
$\ket{g}$ states is largely reduced and has a characteristic
alternating behavior.
A more insightful understanding is obtained by looking at the
temperature evolution of $C^{(S)}_\delta$ for the lattice Kondo problem,
reported in \figu{figSM4}(d).
The spin correlation is initially concentrated at the impurity site
for large temperatures. Upon cooling down the system, the correlations
start to spread through the lattice until reaching an alternating
profile, which, however, does not change sign.
Notably, the local correlation at the impurity site ($\delta=0$) is
essentially temperature independent.

%


\end{document}